# Electron-phonon interactions in LuH$_2$, LuH$_3$, and LuN

Tenglong Lu[1,2], Sheng Meng[1,2,3]*, Miao Liu[1,3,4]*

[1]*Beijing National Laboratory for Condensed Matter Physics, Institute of Physics, Chinese Academy of Sciences, Beijing 100190, China*

[2]*School of Physical Sciences, University of Chinese Academy of Sciences, Beijing 100190, China*

[3]*Songshan Lake Materials Laboratory, Dongguan, Guangdong 523808, China*

[4]*Center of Materials Science and Optoelectronics Engineering, University of Chinese Academy of Sciences, Beijing 100049, China*

*Corresponding author: smeng@iphy.ac.cn, mliu@iphy.ac.cn*

## Abstract

This paper presents the calculation results of electron-phonon interactions within the LuH$_2$, LuH$_3$, and LuN systems under 0 GPa and 10 GPa via density functional theory at the GGA-PBE level. The purpose of this work is to provide useful data that may be of the interests of the superconducting community as it was reported that the Lu-H-N compound is likely to be a room-temperature superconductor under 1 GPa [Nature, 615, 244 (2023)].

## Introduction

Recently, scientists from university of Rochester declared that LuH$_{3-\delta}$N$_\epsilon$ compound is superconductive at 21°C under 1 GPa.[1] The compound according to their characterization and analysis might be a mixed phase of LuH$_{3-\delta}$N$_\epsilon$ and LuN$_{1-\delta}$H$_\epsilon$, but not firmly determined. This observation, if valid, would be a historic discovery of the condensed matter physics, hence promoting the community to prove or disprove the experiments.[2-6]

Previously, employing the high-throughput density functional theory (DFT) as well as the in-house world-class database, which is atomly.net, we have demonstrated that there are no thermodynamically stable ternary phases within the Lu-H-N system under the pressure up to 10 GPa, and the X-ray diffraction characterization of the experimental sample is highly likely to be the mixed phase of $LuH_2$ and LuN.[7] Several follow-up studies confirm that the $LuH_2$ changes its color from blue to bright red when external pressure changes from 1 atm to 4 GPa,[8, 9] is in very good agreement with the experimental result in Ref. [1]. All of those existing studies suggest that $LuH_2$, $LuH_3$, and LuN are the compounds of interest and the derived phase should be the doped phase of them.[10-13] Hence, this work presents the calculated results of electron-phonon interactions within the $LuH_2$, $LuH_3$, and LuN systems under 0 GPa and 10 GPa, which hopefully would provide a useful reference for the community.

## Methods

The structural optimization, phonon dispersions, electron-phonon coupling (EPC), and superconducting properties were carried out under 0 and 10 GPa with the QUANTUM ESPRESSO (QE) package.[14] The plane-wave kinetic-energy cutoff and the energy cutoff for charge density were set as 100 and 400 Ry, respectively. A Methfessel-Paxton smearing width of 0.02 Ry was adopted to calculate the self-consistent electron density. A Brillouin zone (BZ) k-point mesh of 12 × 12 × 12 was adopted for $LuH_2$ and LuN, and a 6 × 6 × 6 BZ k-point mesh was chosen for $LuH_3$ instead. The dynamic and EPC matrix elements were computed with an 8 × 8 × 8 q mesh. The density functional perturbation (DFPT) and Eliashberg theories were used to quantify the phonon and EPC properties.[15, 16]

## Results

Previously, scientists found that the $LaH_{10}$ and $YH_9$ are BCS superconductors thus their superconductivity can be roughly derived from the Eliashberg theory.[16-20] If the Lu-H-N compound is a BCS superconductor, then we can conduct the same treatment by calculating their electron-phonon interactions from DFT too. Figure 1 exhibits the electron-phonon interactions within the $LuH_2$, $LuH_3$, and LuN systems under 0 GPa and 10 GPa based on the DFT calculations. Table 1 presents some numbers from the calculations. Here, we set the empirical parameter μ*, which represents the effective screened Coulomb repulsion, as 0.1 for all the derivations, otherwise stated.

According to our calculation, $LuH_2$ becomes superconducting at very low temperatures ($T_c$ = 0.026 K), in good agreement with Ref. [12]. The hydrostatic pressure of 10 GPa cannot help to shoot up the $T_c$. LuN is not a superconductor under 0 GPa and 10 GPa. LuH$_3$ ($P\bar{3}c1$) is not dynamically stable according to the phonon calculation. The LuH$_3$ ($Fm\bar{3}m$) is thermodynamically unstable based on our evaluation as the energy above the hull ($E_{hull}$) is as high as 91 meV/atom, and is dynamically unstable according to Ref. [1].

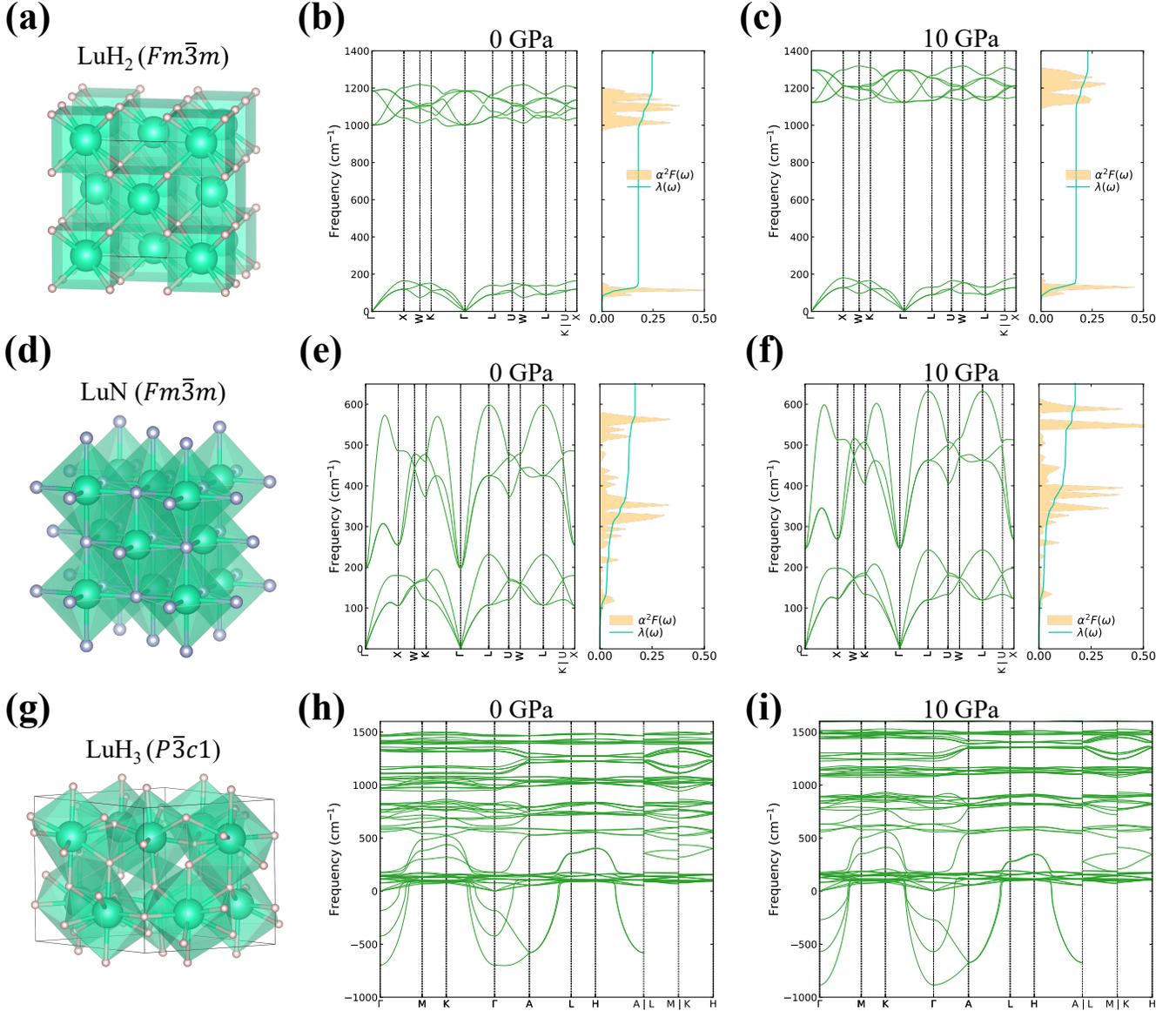

**Figure 1.** Calculated electron-phonon interactions of (b) LuH$_2$ at 0 GPa, (c) LuH$_2$ at 10 GPa, (e) LuN at 0 GPa, (f) LuN at 10 GPa, (h) LuH$_3$ at 0 GPa, and (i) LuH$_3$ at 10 GPa. Detailed crystal structures of (a) LuH$_2$, (d) LuN, and (g) LuH$_3$ were shown in left panels.

**Table 1.** The electron phonon coupling constant λ, logarithmic average frequency $\omega_{ln}$, and superconducting critical temperature $T_c$ for $LuH_2$, LuN, and $LuH_3$ under 0 and 10 GPa. For $LuH_3$, despite its thermodynamic stability, we predicted it was dynamically unstable under both 0 and 10 GPa. Hence, we did not calculate its superconducting properties here.

| Formula | IDs Atomly[21] | IDs ICSD[22] | Pressure (GPa) | λ | $\omega_{ln}$(K) | $T_c$(K) | Is superconductor |
|---|---|---|---|---|---|---|---|
| $LuH_2$ | 1000314722 | 56067 | 0 | 0.259 | 300 | 0.026 | Yes |
|  |  |  | 10 | 0.242 | 315 | 0.010 | Yes |
| LuN | 3001350567 | 44779 | 0 | 0.179 | 410 | 0.000 | No |
|  |  |  | 10 | 0.167 | 497 | 0.000 | No |
| $LuH_3$ | 0000079762 | 638277 | 0 | – | – | – | – |
|  |  |  | 10 | – | – | – | – |

## Conclusions

The DFT calculation suggests that the pure phases of $LuH_2$, $LuH_3$, and LuN are not capable of hosting the room temperature superconductivity within the BCS theoretical framework.